\documentclass[fleqn,usenatbib]{mnras}

\usepackage[T1]{fontenc}
\usepackage{ae,aecompl}

\usepackage{graphicx}	
\usepackage{amsmath}	
\usepackage{amssymb}	

\usepackage[flushleft]{threeparttable}
\usepackage{booktabs}
\usepackage[british]{babel}
\usepackage{newtxtext}
\usepackage[frenchmath,varg]{newtxmath}

\title[Polarisation of the optically-thin jet in M87]{Low optical polarisation at the core of the optically-thin jet of M87}

\author[A. Yrupe Fresco et al.]{
A.Y. Fresco$^{1,2,4}$\thanks{E-mail: \href{mailto:afresco@mpe.mpg.de}{\sf afresco@mpe.mpg.de}},
J.A. Fern\'andez-Ontiveros$^{3,4,5}$,
M.A. Prieto$^{4,5}$,
J.A. Acosta-Pulido$^{4,5}$,
\newauthor
A. Merloni$^{1}$\\
\\
$^{1}$Max-Planck-Institut f\"ur Extraterrestrische Physik (MPE), Giessenbachstrasse 1, D--85748 Garching, Germany\\
$^{2}$Universitat de Barcelona, Gran Via de les Corts Catalanes 585, E--08007 Barcelona, Spain\\
$^{3}$Istituto di Astrofisica e Planetologia Spaziali (INAF--IAPS), Via Fosso del Cavaliere 100, I--00133 Roma, Italy\\
$^{4}$Instituto de Astrof\'isica de Canarias (IAC), C/V\'ia L\'actea s/n, E--38205 La Laguna, Tenerife, Spain\\
$^{5}$Universidad de La Laguna (ULL), Dpto. Astrof\'isica, Avd. Astrof\'isico Fco. S\'anchez s/n, E--38206 La Laguna, Tenerife, Spain\\
}

\date{\today}

\pubyear{2020}

\begin{document}
\label{firstpage}
\pagerange{\pageref{firstpage}--\pageref{lastpage}}
\maketitle

\begin{abstract}
We study the optical linear and circular polarisation in the optically-thin regime of the core and jet of M87. Observations were acquired two days before the Event Horizon Telescope (EHT) campaign in early April 2017. A high degree ($\sim 20$ per cent) of linear polarisation (\textit{P}$_{\rm lin}$) is detected in the bright jet knots resolved at $\sim 10\, \rm{arcsec}$ to $23\, \rm{arcsec}$ ($0.8$--$1.8\, \rm{kpc}$) from the centre, whereas the nucleus and inner jet show \textit{P}$_{\rm lin} \lesssim 5$ per cent. The position angle of the linear polarisation shifts by $\sim 90\degr$ from each knot to the adjacent ones, with the core angle perpendicular to the first knot. The nucleus was in a low level of activity (\textit{P}$_{\rm lin} \sim 2$--$3$ per cent), and no emission was detected from HST-1. No circular polarisation was detected either in the nucleus or the jet above a $3\sigma$ level of \textit{P}$_{\rm circ} \leq 1.5$ per cent, discarding the conversion of \textit{P}$_{\rm lin}$ into \textit{P}$_{\rm circ}$. A disordered magnetic field configuration or a mix of unresolved knots polarised along axes with different orientations could explain the low \textit{P}$_{\rm lin}$. The latter implies a smaller size of the core knots, in line with current interferometric observations. Polarimetry with EHT can probe this scenario in the future. A steep increase of both \textit{P}$_{\rm lin}$ and \textit{P}$_{\rm circ}$ with increasing frequency is expected for the optically-thin domain, above the turnover point. This work describes the methodology to recover the four Stokes parameters using a $\lambda/4$ wave-plate polarimeter.

\end{abstract}




\section{Introduction}

The majority of active galactic nuclei (AGN) in the local Universe remain in a dim activity state characterised by a low accretion rate and a modest luminosity, in contrast with bright quasars and Seyfert nuclei. These are known as Low-Luminosity AGN (LLAGN), typically with bolometric luminosities below $10^{42}\, \rm{erg\,s^{-1}}$, and represent a significant fraction ($\sim$ 1/3) of all nearby galaxies \citep{2008ARA&A..46..475H}. The spectral energy distribution (SED) of bright AGN such as quasars and Seyfert nuclei is dominated mainly by thermal processes, i.e. the accretion disc in the optical/ultraviolet (UV) and the central reprocessed dust emission in the infrared (IR), which obscures the central engine \citep{1999PASP..111....1K}. In contrast, non-thermal processes seem to be common in LLAGN, usually associated with jets \citep[e.g.][]{2016MNRAS.457.3801P,2007ApJ...663..808P,2009ApJ...705..356B,2008ApJ...681..905M,2019MNRAS.485.5377F} and/or radiatively inefficient accretion flows \citep[e.g.][]{1995ApJ...452..710N,2008ARA&A..46..475H}.

AGN jets are complex environments involving high-energy physics, where both particle acceleration and cooling take place. The continuum emission is mainly powered by synchrotron self-absorbed radiation \citep{1979ApJ...232...34B}, typically peaking in the radio to millimetre range, plus inverse Compton cooling in the X-rays \citep[e.g.][]{2013MNRAS.435.2520G}. Furthermore, jets represent a strong source of feedback for their host galaxies, e.g. by drilling the interstellar medium (ISM), inflating cavities, and heating the circumgalactic medium (e.g. \citealt{2012ARA&A..50..455F}, and references therein). Polarimetry is an ideal tool to probe the inner structure of jets, as the non-thermal emission is typically strongly polarised. In particular, the highest degree of both linear and circular polarisation is expected for the optically-thin synchrotron continuum \citep{1970ranp.book.....P,1977OISNP..89.....P}, typically found in the millimetre to UV range. However, the majority of AGN show a relatively low degree of linear polarisation $\lesssim 20$ per cent in the optical range, mostly due to the contribution of non-polarised light from the stars in the host galaxy and inhomogeneities in the magnetic field structure. The latter produce radiation with different polarisation angles and thus decrease the net degree of polarisation measured if they are not spatially resolved by the observations. Few studies include optical measurements of the Stokes V parameter for the circular polarisation \citep[e.g.][]{2010A&A...520L...7H}, mainly due to technical reasons (see Section\,\ref{obs}), and most of them are dedicated to blazars, which are strongly beamed AGN.

One of the most studied jets in the local Universe is found in the nucleus of M87, a prototypical LLAGN ($D = 16.4\, \rm{Mpc}$; \mbox{\citealt{2005ApJ...634.1002J}}). This galaxy harbours one of the most prominent optical jets known, extensively studied over the last decades. Previous studies of the linear polarisation in the optical range have shown that the core of the jet has a low degree of linear polarisation ($\lesssim 10$ per cent) whereas that of the knots along the jet is higher \citep[$10$--$30$ per cent;][]{1989A&A...224...17F,1999AJ....117.2185P,2016ApJ...832....3A}. The cause of this difference in polarisation between the core and the rest of the jet is not clear, a possible explanation includes the presence of dust in the nucleus which could obscure and block the polarisation signal. Yet, there is no sign of dust in the nucleus of M87 as confirmed by high angular resolution dust maps \citep{2010A&A...518L..53B,2016MNRAS.457.3801P}. Another possibility is that the magnetic field in the region of the nucleus could be disordered, and thus the measured polarisation would include a mix of different signals. The presence of extended ionised gas over the nuclear region \citep{1994ApJ...435L..35H} has proved to cause Faraday rotation in the millimetre/sub-millimetre continuum, changing the position angle of the linear polarisation \citep{2014ApJ...783L..33K}. However, to our knowledge no Faraday depolarisation has been reported so far for extragalactic jets in the optical range.

In this work we investigate the polarised emission from the nucleus and jet of M87 using full Stokes imaging in the optical range. This allows us to characterise the distribution of both the linear and the circular polarisation for the optically-thin part of the jet continuum. In particular we aim to understand the low degree of optical linear polarisation in the core of M87, as well as the presence of possible circular polarisation. As a byproduct of the study, we provide the complete methodology followed to recover the four Stokes parameters using a polariser including a $\lambda/4$ waveplate instead of the common $\lambda/2$ waveplate configuration, which does not allow to measure the Stokes V parameter. This study is organised as follows: in Section\,\ref{obs} we describe the observations, the detailed methodology is explained in Section\,\ref{method}, in Section\,\ref{results} we present the main results of this work, which are discussed in Section\,\ref{discuss}. Finally, a summary of the present study and the main conclusions can be found in Section\,\ref{sum}.


\section{Observations and Data Reduction}\label{obs}

\begin{figure}
\centering
\includegraphics[width=\columnwidth]{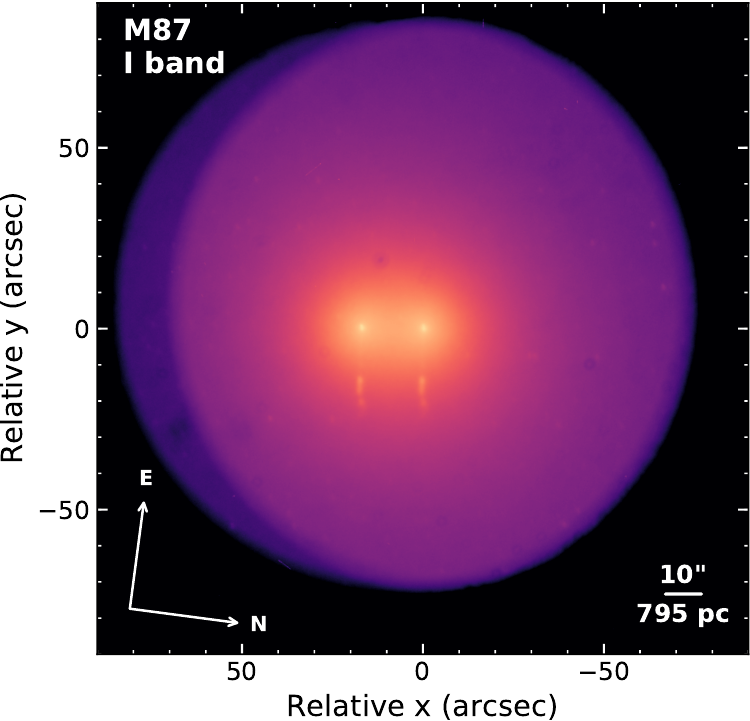}
\caption{Raw image showing the nucleus and optical jet of M87 in the \textit{I}~band. The field vignetting is caused by the limited size of the calcite crystal ($\sim 140\, \rm{arcsec}$). The moon-shaped shadow on the left side of the image corresponds to the area illuminated only by the extraordinary beam, while both beams overlap in the rest of the illuminated area. Note that a partial superposition of the host galaxy unpolarised light occurs for the extraordinary image (left) and the ordinary image (right). Both rays are separated by an angle of $6\fdg2$ ($15$ arcsec). Dark spots corresponds to a lower transmission ($\sim 30$ per cent less on the darkest spots) caused by impurities in the calcite crystal. M87 was placed in a region free of low transmission spots. The reference system is centred on the position of the nucleus in the ordinary beam.}\label{fig_raw}
\end{figure}

The polarimetric observations were acquired with the Alhambra Faint Object Spectrograph and Camera (ALFOSC) instrument in the Nordic Optical Telescope (NOT), at the Roque de los Muchachos Observatory (La Palma, Spain). This is one of the few facilities with the capability to measure the full Stokes parameters. While the majority of astronomical polarimeters introduce a $\lambda / 2$ waveplate and a birefringent element along the light path to measure the degree of linear polarisation, the polarimetric unit in ALFOSC (FAPOL) includes a $\lambda / 4$ waveplate to allow the measurement of the circular polarisation. The birefringent element in FAPOL consists on a calcite crystal that provides two orthogonal polarised beams, i.e. phase-shifted by $90\degr$. Imaging polarimetry for M87 was acquired for two broad-band filters named \texttt{v\_Bes530\_80} (hereafter \textit{V}~band), $\lambda_c = 530\, {\textrm{nm}}$, $\Delta \lambda = 80\, \rm{nm}$) and \texttt{i\_int797\_157} (\textit{I}~band, $\lambda_c = 797\, {\textrm{nm}}$, $\Delta \lambda = 157\, \rm{nm}$), using 16 different positions of the $\lambda/4$ plate spaced by $22\fdg5$ (see Section\,\ref{method}). The pixel scale of the ALFOSC detector is $0.25\, \rm{arcsec/px}$. The total observing time was of 5 hours, undertaken during the night of April 3rd, 2017, which is only a few days before the Event Horizon Telescope campaign from April 5th to 11th. This event provided the first direct evidence of the shadow formed due to the gravitational bending of light close to the supermassive black hole at the centre of M87 \citep{2019ApJ...875L...1E}. Some clouds were present occasionally during the night but they did not affect the observations of M87. Overall the seeing was low with values in the $0.6$--$1.0\, \rm{arcsec}$ range during the night. Bias, sky- and dome-flat images were taken at the beginning of the night. In order to calibrate the instrumental polarisation, five standard stars were observed during the night. Two of them are linearly polarised, two are zero-polarisation standards, and one has circular polarisation. Catalogue values for the standards are provided in Table\,\ref{tab_std}.

This research made use of \texttt{ccdproc}, an Astropy package for
image reduction \citep{2015ascl.soft10007C} programmed in \textsc{python}, and our own scripts developed using the \textsc{spyder}\footnote{\url{https://docs.spyder-ide.org}} environment. The bias images were combined using the median and subtracted from the rest of the images. Variations in the pixel-to-pixel sensitivity were corrected using the dome-flats for the \textit{V}~band and sky flats for the \textit{I}~band. Dome flats are preferred over the sky flats because the dome illumination does not introduce a significant polarisation, and thus the flats can be acquired with the calcite inserted along the optical path to correct for the calcite transmission. The latter introduce a number of dark spots in the image (see Fig.\,\ref{fig_raw}). During the acquisition of the sky flats the calcite was removed due to the polarisation of the sky background. Unfortunately the exposure times of the \textit{I}~band dome flats were overestimated, and thus the counts reached the non-linear range response of the detector. Therefore we used the sky flats to correct the pixel-to-pixel response of the detector in the \textit{I}~band. Using the sky flats for the \textit{V}~band does not alter the results presented in this work. The polarisation of the standard stars was estimated from the aperture fluxes, subtracting the sky background contribution estimated in a ring around the star. Note that GD\,319 is a double star, thus the aperture used was larger to include the two stars, since the catalogue value corresponds to the integrated light of the system. In Section\,\ref{method} we describe the methodology followed to recover the full Stokes parameters from the modulated intensities detected by ALFOSC/FAPOL. The polarisation values obtained for the standards are shown in Table\,\ref{tab_std}.
\begin{table*} 
 \begin{center}
   \caption{Reference values for the polarisation degree were taken from \citet{1990AJ.....99.1243T} for the non-polarised standars, \citet{1992AJ....104.1563S} for the linearly polarised standars, and \citet{1975ApJ...196..819L} for the circularly polarised star.}\label{tab_std}
 \begin{tabular}{cccccccc} 
 \hline
 \bf Name & \bf Type & \bf Ref. filter & \bf Ref. $P$  & \bf Ref. $\chi$ & \bf Obs. filter & \bf Obs. $P$ & \bf Obs. $\chi$ \\ 
      &           &       & (per cent) & (deg) & & (per cent) & (deg) \\[0.5ex] 
 \hline
 Grw\,+70\degr8247 & Circular & - & \textit{P}$_{\rm circ} = 3.20 \pm 0.07$ & - & I & \textit{P}$_{\rm circ} = 3.0 \pm 0.5$ & - \\
 GD\,319 & None & B & \textit{P}$_{\rm lin} = 0.040 \pm 0.047$  & - & V & \textit{P}$_{\rm lin} = 0.1 \pm 0.5$ & -\\
 BD+33\,2642 & None & B & \textit{P}$_{\rm lin} = 0.145 \pm 0.029$ & - & V & \textit{P}$_{\rm lin} = 0.3 \pm 0.5$ & - \\
 VI\,Cyg\,\#12 & Linear & V & \textit{P}$_{\rm lin} = 8.95 \pm 0.09$ & $115.03 \pm 0.28$ & V & \textit{P}$_{\rm lin} = 9.8 \pm 0.5$ & $96.8$ \\
 Hiltner\,960 & Linear & V & \textit{P}$_{\rm lin} = 5.66 \pm 0.03$ & $54.79 \pm 0.11$ & V & \textit{P}$_{\rm lin} = 5.8 \pm 0.5$ & $39.2$ \\
 \hline
\end{tabular}
\end{center}
\end{table*} 

\begin{figure}
\centering
\includegraphics[width=\columnwidth]{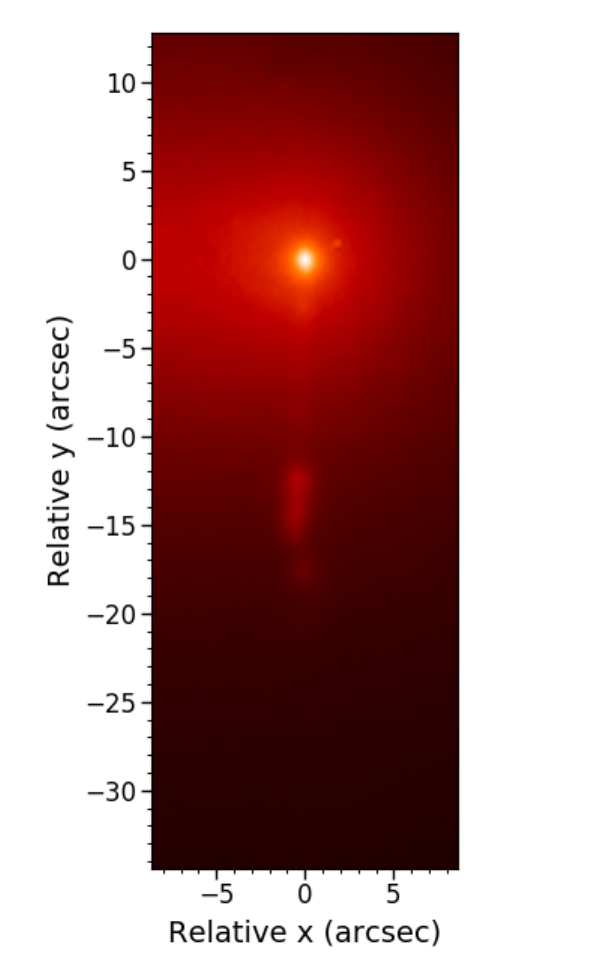}
\caption{Trimmed image of M87 corresponding to the extraordinary ray in the I band.}\label{trimmed}
\end{figure}

Prior to the polarimetric analysis, the images of M87 were registered and combined. In order to avoid superposition between the ordinary and extraordinary images of the jet, the latter was oriented along the vertical axis by rotating the instrument at the beginning of the night (see Fig.\,\ref{fig_raw}). The alignment of the individual frames was based on the centroid measured at the two positions of the point-like nucleus (ordinary and extraordinary). The average nuclear position was then used to shift and combine the frames, to obtain a final image per filter and for each of the 16 different positions of the $\lambda/4$ waveplate, which was then trimmed to separate the ordinary and extraordinary images. Finally, the diffuse emission from the host galaxy was removed in order to avoid the dilution of the polarised signal in the nucleus by the contribution of the underlying stellar light. This was achieved by applying a low frequency filtering in the Fourier space \citep[e.g.][]{2019MNRAS.488.2800H}, i.e. by removing the low spatial frequencies associated with the extended emission. To perform this task in the least intrusive way we first obtained a total intensity image from the different exposures at $0\degr$, $90\degr$, $180\degr$, and $270\degr$ (see Table\,\ref{tab_mueller}), which was then filtered in the Fourier space to keep only $8$ per cent of the lowest spatial frequencies. The fraction of spatial frequencies adopted was optimised after an iterative process to clean as much background light as possible while keeping the residuals in the cleaned images below the noise level, once the diffuse light is subtracted. The reconstructed image after the Fourier filtering corresponds to the intensity of the extended starlight contribution, and thus was subtracted from all the individual frames prior to the polarimetric analysis in Section\,\ref{method}. Hereafter we consider the nucleus to be the reference of the relative coordinates ($x$, $y$), where $y$ corresponds to the distance along the jet direction (see Fig.\,\ref{trimmed}). The Stokes parameters maps and the polarisation maps were obtained using the same method applied to the polarimetric standards (Section\,\ref{method}).


\section{Methodology}\label{method}

The polarisation state of the electromagnetic radiation can be described using the four Stokes parameters. $\textit{I}$ stands for the total intensity (after background starlight removal), \textit{Q} and \textit{U} describe the linear polarisation component, and \textit{V} correspond to the circular polarisation. In this Section we elaborate in detail the method applied to recover the four Stokes parameters from the modulated intensity registered by ALFOSC/FAPOL. The drawback of using a $\lambda / 4$ waveplate --\,required to measure the circular polarisation\,-- is a more complex methodology to derive \textit{Q} and \textit{U}, when compared to the method using a $\lambda / 2$ waveplate. Although the latter is available in ALFOSC, the retarder cannot be changed during the observing night, thus we derive here the equations required to recover the linear polarisation using the same $\lambda/4$ waveplate used to measure the circular polarisation.

To recover the Stokes parameters for the standard stars and M87, we applied the M\"uller calculus to the optical elements in FAPOL. FAPOL transforms the \textit{Q}, \textit{U}, and \textit{V} parameters into a modulated intensity (\textit{I'}), which can be then recorded by the CCD detector. Following \citet{afa12}, the polarisation state of the electromagnetic radiation transmitted through a phase retarder and a linear polariser can be described by:
\begin{equation} \label{eq:lambda4}
\begin{bmatrix}
I' \\
Q' \\
U' \\
V' 
\end{bmatrix} 
= \frac{1}{2} M
\times W_{\lambda/4}
\times  \begin{bmatrix}
I \\
Q \\
U \\
V 
\end{bmatrix}
\end{equation}
where $M$ corresponds to the calcite crystal, and $W_{\lambda / 4}$ represents the $\lambda / 4$ waveplate\footnote{We note a possible typo in eq.\,10 of \citet{afa12}, where the last element of the $W_{\lambda / 4}$ matrix is $-1$ instead of $0$.}:
\begin{align}
  \begin{split}
  M &=
  \begin{bmatrix}
    1 & C & S & 0 \\
    C & C^2 & CS & 0 \\
    S & CS & S^2 & 0 \\
    0 & 0 & 0 & 0
  \end{bmatrix}\\
  W_{\lambda / 4} &=
  \begin{bmatrix}
    1 & 0 & 0 & 0 \\
    0 & C^2_{\lambda}  & C_{\lambda}S_{\lambda} & -S_{\lambda} \\
    0 & C_{\lambda}S_{\lambda} &  S^2_{\lambda} & C_{\lambda} \\
    0 & S_{\lambda} & -C_{\lambda} & 0
  \end{bmatrix}\\
  \end{split}
  \end{align}
where 
\begin{align*}
  &C = \cos(2\varphi), S = \sin(2\varphi)\\
  &C_{\lambda} = \cos(2\theta) = \cos(2\theta' + 90\degr)\\
  &S_{\lambda} = \sin(2\theta) = \sin(2\theta' + 90\degr)
\end{align*}
$\varphi$ is the position angle of the polariser transmission axis, and $\theta$ is the orientation of the retarder fast axis. The calcite $M$ acts as a double polariser producing two linearly polarised light beams. The extraordinary beam emerges parallel to the calcite fast axis [$M(\varphi = 0\degr)$], which was fixed during the observations, while the ordinary beam is polarised in the orthogonal direction [$M(\varphi = 90\degr)$]. Since the detector is sensitive to the intensity \textit{I'} in Eq.\,\ref{eq:lambda4}, the four Stokes parameters can be recovered when \textit{I'} is modulated as a function of $\theta$, that is observing with the $\lambda / 4$ waveplate oriented along 8 different angles spaced by $22\fdg5$. In order to cancel the effect caused by the different transmission along the optical light path of the ordinary and the extraordinary beams, and also increase the signal-to-noise (\textit{S/N}) ratio of the measurement, we used $16$ different angles from $\theta = 0\degr$ to $337\fdg5$ in steps of $22\fdg5$. Following the notation used by the acquisition software at the telescope, we define $\theta' = \theta - 45\degr$, where $\theta' = 0$ corresponds to an offset of $45\degr$ between the fast axes of the calcite and the retarder. Thus, for each $\theta$ angle of the phase plate, the image acquired contains two intensity measurements corresponding to the extraordinary (\textit{I'}$_{\rm e}$ at $\varphi = 0\degr$) and the ordinary beams ($I'_{\rm o}$ at $\varphi = 90\degr$). Therefore, the modulated intensities registered by the detector are:
\begin{align}\label{eq_Ieo}
  \begin{split}
  I'_{\rm e}(\theta') &= \frac{1}{2}\left(I + Q C^2_{\lambda} + U C_{\lambda} S_{\lambda} - V S_{\lambda}\right)\\
  I'_{\rm o}(\theta') &= \frac{1}{2}\left(I - Q C^2_{\lambda} - U C_{\lambda} S_{\lambda} + V S_{\lambda}\right)
  \end{split}
\end{align}

Considering the 8 independent positions of the $\lambda / 4$ waveplate and the two linearly polarised beams, we obtain a system of 16 linear equations for each filter that can be solved to recover the full Stokes parameters (see Table\,\ref{tab_mueller}).

For instance, the parameters \textit{V} and \textit{Q} can be obtained directly from the ordinary and extraordinary beams at $\theta' = 0\degr$ and $45\degr$, respectively:

\begin{equation}\label{eq:polcir}
  \frac{V}{I} = \frac{I'_{\rm o}(\theta' = 0\degr) - I'_{\rm e}(\theta' = 0\degr)}{I'_{\rm o} (\theta' = 0\degr) + I'_{\rm e}(\theta' = 0\degr)};\\
  \frac{Q}{I} = \frac{I'^{45\degr}_{\rm e} - I'^{45\degr}_{\rm o}}{I'^{45\degr}_{\rm e} + I'^{45\degr}_{\rm o}}
\end{equation}

Note that the total intensity \textit{I} includes both the jet emission and the unpolarised stellar component. The latter is not contributing to \textit{V} Stokes but decreases the \textit{V/I} ratio, and therefore a correction for the stellar unpolarised emission is required to measure the intrinsic polarisation of the jet components. For this purpose, we obtained a total intensity map using the images of the first and fifth rows in Table\,\ref{tab_mueller}, that is combining the ordinary and extraordinary beams acquired with $\theta = 0\degr, 180\degr, 90\degr,$ and $270\degr$. Then a Fourier spatial filter was applied to \textit{I} in order to remove the small scale emission associated with the jet emission. The resulting image \textit{I}$^*$ was used to subtract the stellar contribution from all the individual \textit{I}$'_{\rm o}(\theta')$ and \textit{I}$'_{\rm e}(\theta')$ frames.

In order to cancel the transmission effects we followed the ``ratio method" \citep[e.g.][]{2009PASP..121..993B}. \textit{R}$_v$ is defined as the ratio of the intensity measurements (\textit{I'}) that are modulated only by the Stokes \textit{V} ($0\degr$ and $90\degr$, $180\degr$ and $270\degr$). Finally, for each pair of measurements we can obtain an independent estimate of \textit{R}$_v$:
\begin{align}
  \begin{split}
  R_v &= \sqrt{\frac{I'^{0\degr}_{\rm o}/I'^{0\degr}_{\rm e}}{I'^{90\degr}_{\rm o}/I'^{90\degr}_{\rm e}}}= \sqrt{\frac{I'^{180\degr}_{\rm o}/I'^{180\degr}_{\rm e}}{I'^{270\degr}_{\rm o}/I'^{270\degr}_{\rm e}}} = \frac{I + V}{I - V}\\
  v &= \frac{V}{I} = \frac{R_v - 1}{R_v + 1}
  \end{split}
\end{align}

The two \textit{R}$_v$ estimates were averaged to derive \textit{v}, i.e. the relative contribution of the Stokes \textit{V} to the total intensity\footnote{The circular polarisation is defined as negative for left-handed polarisation (clockwise rotation of the polarised vector), or positive for right-handed polarisation (counter-clockwise rotation).}. On the other hand, the measurement of the linear polarisation using a $\lambda / 4$ waveplate instead of a $\lambda / 2$ requires a more elaborated algebra. Note that the contribution of \textit{Q} to the ordinary and extraordinary beams in Eq.\,\ref{eq_Ieo} does not change its sign with $\theta'$, which hampers the cancellation of the transmission factor between orthogonal orientations of the $\lambda / 4$ waveplate. This can be solved by introducing the ratio \textit{R}$_q$, which cancels the transmission factor for the Stokes \textit{Q} taking advantage of the \textit{R}$_v$ ratio previously defined. This provides four independent estimates for \textit{R}$_q$, that were averaged to obtain \textit{q}:
\begin{align}\label{eq_Rq}
  \begin{split}
  R_q &= \frac{1}{R_v}\frac{I'^{45\degr}_{\rm e}/I'^{45\degr}_{\rm o}}{I'^{0\degr}_{\rm e}/I'^{0\degr}_{\rm o}} = \frac{1}{R_v}\frac{I'^{225\degr}_{\rm e}/I'^{225\degr}_{\rm o}}{I'^{180\degr}_{\rm e}/I'^{180\degr}_{\rm o}}\\
  &= R_v \frac{I'^{135\degr}_{\rm e}/I'^{135\degr}_{\rm o}}{I'^{90\degr}_{\rm e}/I'^{90\degr}_{\rm o}} = R_v \frac{I'^{315\degr}_{\rm e}/I'^{315\degr}_{\rm o}}{I'^{270\degr}_{\rm e}/I'^{270\degr}_{\rm o}} = \frac{I + Q}{I - Q}\\
  q &= \frac{Q}{I} = \frac{R_q - 1}{R_q + 1}
  \end{split}
\end{align}

Since \textit{U} does not modulate \textit{I'} without the contribution of \textit{Q} and \textit{V}, a system of equations including all the Stokes parameters must be solved. Similarly as in Eq.\,\ref{eq_Rq}, we define \textit{R}$_{u1}$, \textit{R}$_{u2}$, \textit{R}$_{u3}$, and \textit{R}$_{u4}$, including also the $R_v$ ratio to cancel the transmission factors between the ordinary and extraordinary beams:
\begin{equation}
  \begin{split}
  R_{u1} &= R_v\frac{I'^{67\fdg5}_{\rm e}/I'^{67\fdg5}_{\rm o}}{I'^{90\degr}_{\rm e}/I'^{90\degr}_{\rm o}} = R_v \frac{I'^{247\fdg5}_{\rm e}/I'^{247\fdg5}_{\rm o}}{I'^{270\degr}_{\rm e}/I'^{270\degr}_{\rm o}}\\
  R_{u2} &= \frac{1}{R_v}\frac{I'^{157\fdg5}_{\rm e}/I'^{157\fdg5}_{\rm o}}{I'^{180\degr}_{\rm e}/I'^{180\degr}_{\rm o}} = \frac{1}{R_v}\frac{I'^{337\fdg5}_{\rm e}/I'^{337\fdg5}_{\rm o}}{I'^{0\degr}_{\rm e}/I'^{0\degr}_{\rm o}}\\
  R_{u3} &= R_v\frac{I'^{112\fdg5}_{\rm e}/I'^{112\fdg5}_{\rm o}}{I'^{90\degr}_{\rm e}/I'^{90\degr}_{\rm o}} = R_v\frac{I'^{292\fdg5}_{\rm e}/I'^{292\fdg5}_{\rm o}}{I'^{270\degr}_{\rm e}/I'^{270\degr}_{\rm o}}\\
  R_{u4} &= \frac{1}{R_v}\frac{I'^{67\fdg5}_{\rm e}/I'^{67\fdg5}_{\rm o}}{I'^{90\degr}_{\rm e}/I'^{90\degr}_{\rm o}} = \frac{1}{R_v}\frac{I'^{247\fdg5}_{\rm e}/I'^{247\fdg5}_{\rm o}}{I'^{270\degr}_{\rm e}/I'^{270\degr}_{\rm o}}\\
  u &= \frac{U}{I}=\frac{R_{u1} - 1}{R_{u1} + 1} - \frac{R_{u3} - 1 }{R_{u3} + 1} = \frac{R_{u2} - 1}{R_{u2} + 1} - \frac{R_{u4} - 1}{R_{u4} + 1}
  \end{split}
\end{equation}

\begin{table}
\begin{center}
\caption{Results from the matrix calculations for each angle of the waveplate ($\varphi$) and each angle of the calcite polariser ($\theta$) for the ordinary (\textit{I}$'^{\theta}_{\rm o}$) and extraordinary beams (\textit{I}$'^{\theta}_{\rm e}$).}\label{tab_mueller}
\begin{tabular}{ccc} 
\hline
\bf $\lambda/4$ fast axis  & \bf $I'^{\theta}_{\rm o}$  & \bf $I'^{\theta}_{\rm e}$ \\
($\theta$) & ($\varphi = 90\degr$) & ($\varphi = 0\degr$)\\[0.2cm]
 \hline
  $0\degr$, $180\degr$ & $\dfrac{1}{2}(I + V)$ & $\dfrac{1}{2}(I - V)$\\[0.3cm] 
 $22\fdg5$, $202\fdg5$ & $\dfrac{1}{2}\left(I - \dfrac{Q}{2} + \dfrac{U}{2} + \dfrac{\sqrt{2}}{2} V\right)$ & $\dfrac{1}{2}\left(I + \dfrac{Q}{2} - \dfrac{U}{2} - \dfrac{\sqrt{2}}{2} V\right)$ \\[0.3cm]
 $45\degr$, $225\degr$ & $\dfrac{1}{2}(I - Q)$ & $\dfrac{1}{2}(I + Q)$  \\[0.3cm]
 $67\fdg5$, $247\fdg5$ & $\dfrac{1}{2}\left(I - \dfrac{Q}{2} - \dfrac{U}{2} - \dfrac{\sqrt{2}}{2} V \right)$ & $\dfrac{1}{2} \left(I + \dfrac{Q}{2} + \dfrac{U}{2} + \dfrac{\sqrt{2}}{2} V \right)$ \\[0.3cm]
 $90\degr$, $270\degr$ & $\dfrac{1}{2}(I - V)$ & $\dfrac{1}{2}(I + V)$\\[0.3cm] 
 $ 112\fdg5$, $292\fdg5$ & $\dfrac{1}{2}\left(I - \dfrac{Q}{2} + \dfrac{U}{2} - \dfrac{\sqrt{2}}{2} V\right)$ & $\dfrac{1}{2}\left(I + \dfrac{Q}{2} - \dfrac{U}{2} + \dfrac{\sqrt{2}}{2} V\right)$ \\[0.3cm]
 $135\degr$, $315\degr$ & $\dfrac{1}{2}(I - Q)$ & $\dfrac{1}{2}(I + Q)$ \\[0.3cm]
$157\fdg5$, $337\fdg5$ & $\dfrac{1}{2}\left(I - \dfrac{Q}{2} - \dfrac{U}{2} + \dfrac{\sqrt{2}}{2} V\right)$ & $\dfrac{1}{2}\left(I + \dfrac{Q}{2} + \dfrac{U}{2} - \dfrac{\sqrt{2}}{2} V\right)$ \\[0.1cm]
 \hline
\end{tabular}
\end{center}
\end{table}

The two independent measurements for each ratio \textit{R}$_u$ were averaged. These ratios provide two estimates for \textit{u}, that were also averaged. Finally, the percentage of the linear and the circular polarisation (\textit{P}$_{\rm lin}$ and \textit{P}$_{\rm circ}$, respectively) and the position angle of the former ($\chi$) can be obtained from:
\begin{align}
\begin{split}
  &P_{\textrm{lin}} = 100 \times \sqrt{q^2 - \sigma^2_q + u^2 - \sigma^2_u }\\
  &P_{\textrm{circ}} = 100 \times v\\
  &\chi = {PA}_{\textrm{sky}} + \frac{1}{2}\arctan\left(\frac{u}{q}\right)
  \end{split}
\end{align}
where $\sigma^2_q$ and $\sigma^2_u$ are the uncertainties of the \textit{Q} and \textit{U} Stokes parameters, respectively. These are subtracted to correct for the positive bias introduced by the noise for those regions with a low \textit{S/N} ratio. \textit{PA}$_{\rm sky}$ is the position angle of the instrument, defined in the counter-clock direction with respect to the North.

The degree of linear and circular polarisation, and the position angle of the linear polarisation, were obtained for the polarimetric standards using the fluxes measured with aperture photometry and the method detailed along this Section (see Table\,\ref{tab_std}). The final error for the standard stars corresponds to the difference between the published values and our measurements, which is larger than the photometric error in all cases, except for the non-polarised standards. For the images of M87 we measured the error associated with the Stokes parameters $\sigma_q$, $\sigma_u$, and $\sigma_v$ in the corresponding maps as the standard deviation within a square region of a few arcseconds in size, close to the galaxy nucleus where no polarimetric signal was detected. The errors obtained for each filter and Stokes parameter are:
\begin{equation}
\begin{split}
V~{\rm band}: \sigma_q &= 0.36,  \quad \sigma_u = 0.34, \quad \sigma_v = 0.5\\
I~{\rm band}: \sigma_q &= 0.36, \quad \sigma_u = 0.34, \quad \sigma_v = 0.5
\end{split}
\end{equation}

To estimate the total uncertainties for the linear polarisation we applied the standard error propagation:
\begin{equation}
\sigma_{\rm lin} = \sqrt{\sigma^2_q + \sigma^2_u}
\end{equation}

The error in the linear polarisation measurements is $\sigma_{\rm lin} = 0.5$ per cent for both the \textit{V} and the \textit{I}~bands. Since the instrumental polarisation is not significant, no additional correction was applied to the polarisation maps obtained for M87.


\section{Results}\label{results}

\begin{figure}
\centering
\includegraphics[width=0.5\columnwidth]{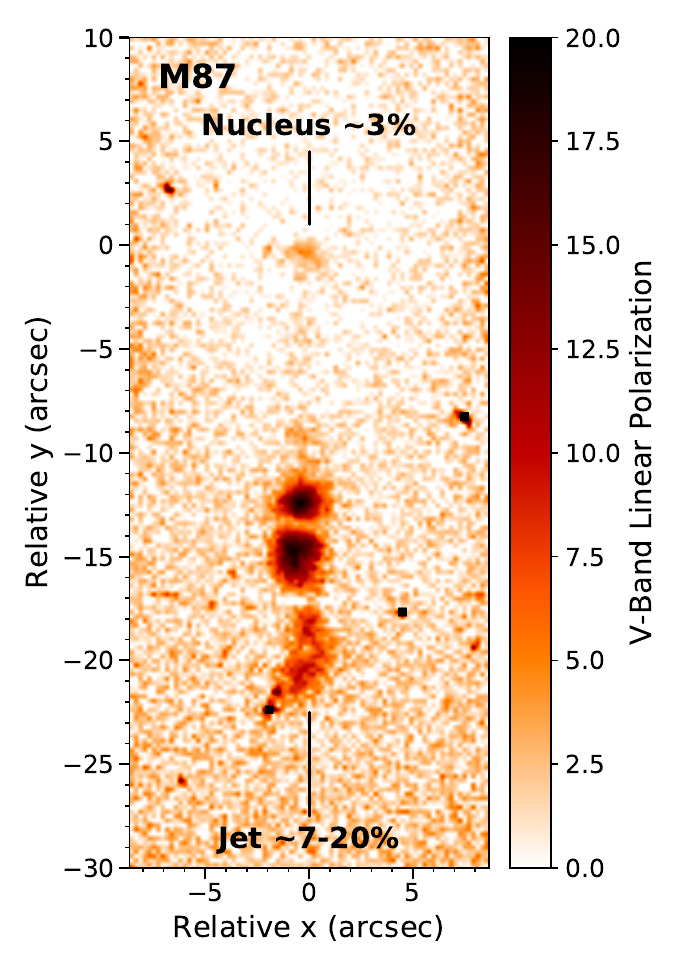}~
\includegraphics[width=0.5\columnwidth]{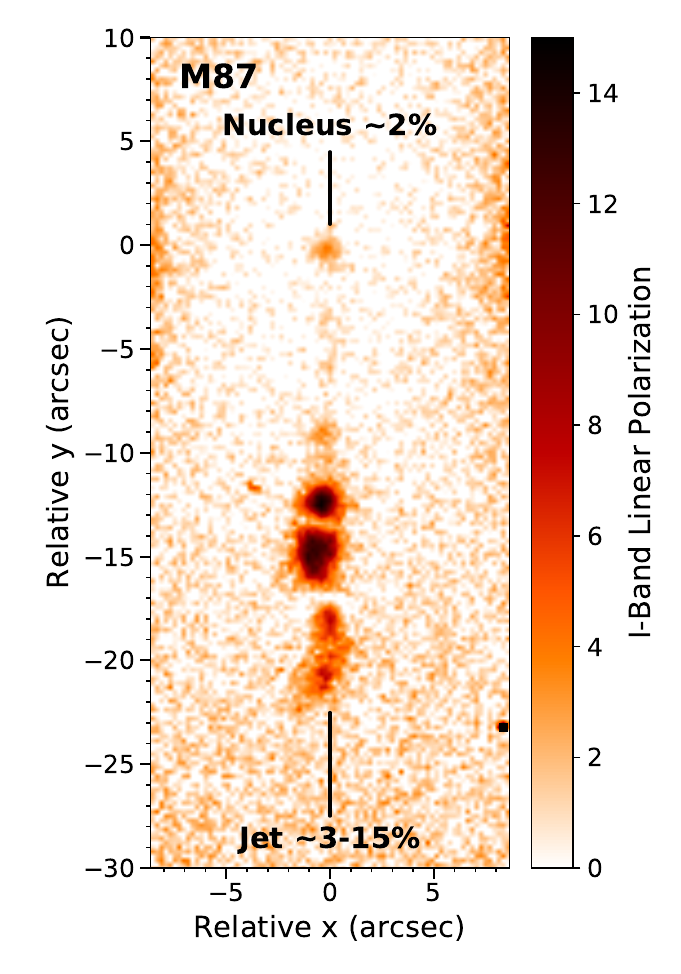}
\caption{Map of the linear polarisation degree in the central region of M87, obtained for the \textit{V} (left) and the \textit{I}~bands (right). The angular coordinates are defined in arcseconds relative to the position of the nucleus.}\label{fig_polmap}
\end{figure}

\begin{figure}
\centering
\includegraphics[width=0.51\columnwidth]{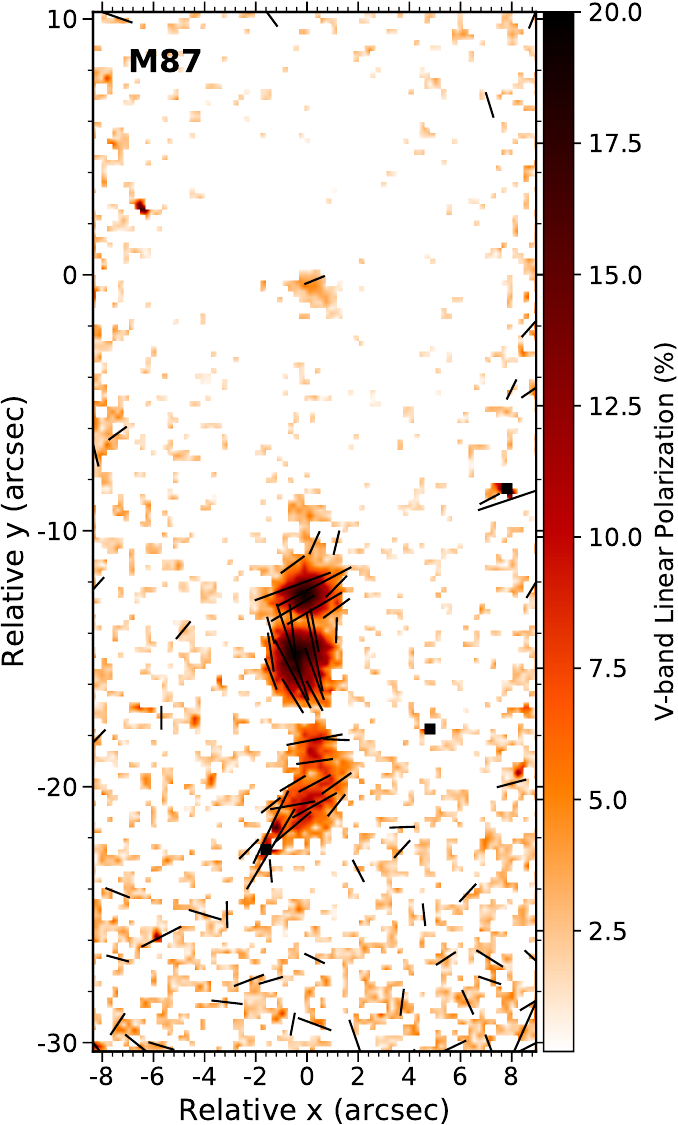}~
\includegraphics[width=0.49\columnwidth]{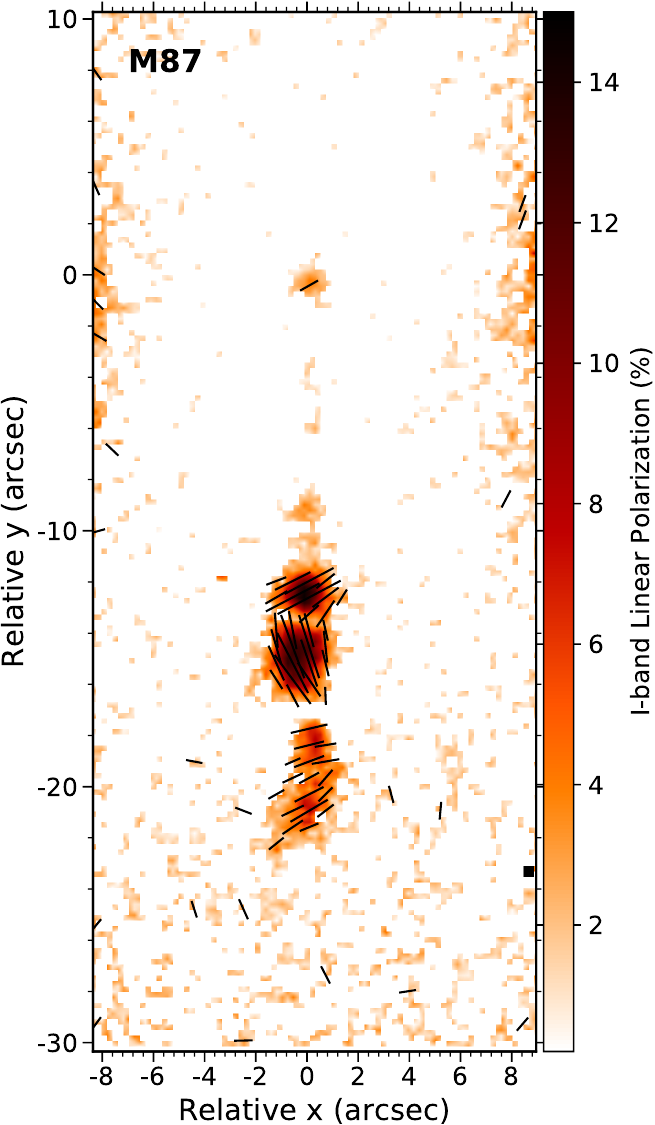}
\caption{Vector map of the linear polarisation angle for the central region of M87, in the \textit{V} (left) and \textit{I}~bands (right). The background images show the linear polarisation maps from Fig.\,\ref{fig_polmap}. The length of each segments in the vector map is proportional to the degree of polarisation in that position.}\label{fig_pa}
\end{figure}

In Fig.\,\ref{fig_polmap} we show the linear polarisation maps obtained for M87 in the \textit{V} (left panel) and \textit{I}~bands (right), using the method described in Section\,\ref{method}. The angle of the linear polarisation is indicated by the vector field in Fig.\,\ref{fig_pa}. The nucleus of M87 shows a mild degree of linear polarisation with $3.1 \pm 0.6$ per cent and $2.3 \pm 0.4$ per cent in the \textit{V} and \textit{I}~bands, respectively, in line with the lowest values previously reported in the literature \citep{2011ApJ...743..119P}. At radio wavelengths the linear polarisation degree of the core remained below $4$ per cent between 2002 and 2009 \citep{2016ApJ...832....3A}, as expected for optically-thick emission in this range \citep{1977OISNP..89.....P}.
The jet knot HST-1 (named by \citealt{1999ApJ...520..621B}), which experienced bright flare events in the past \citep{2009ApJ...699..305H} was in a low state in April 2017 \citep{2019ApJ...879....8S} and thus it was not detected in our maps. Note that the separation from the nucleus to HST-1 is $0.86\, \rm{arcsec}$, whereas the angular resolution of our images is $\sim 0.6\, \rm{arcsec}$.

Following the jet along the $y$ axis in Fig.\,\ref{fig_polmap}, there is a drop in both the total intensity and the linear polarisation to values of $\sim 1$--$2$ per cent, up to $11\, \rm{arcsec}$ away from the centre ($\sim 0.9\, \rm{kpc}$ projected distance). This region covers the knots D, E, F, and I resolved in the \textit{HST} maps by \citet{2016ApJ...832....3A}. From $\sim 0.9\, \rm{kpc}$ on there is an abrupt increment in both the flux intensity (see Fig.\,\ref{trimmed}) and the linear polarisation degree up to $\sim 19$ per cent ($\sim 14$ per cent) in the \textit{V}~band (\textit{I}~band), remaining at this level up to the end of the optical jet at a projected distance of $23\, \rm{arcsec}$ from the nucleus ($1.8\, \rm{kpc}$). The four main knots in the jet are located at $10.9\, \rm{arcsec}$ ($0.9\, \rm{kpc}$), $15\, \rm{arcsec}$ ($1.2\, \rm{kpc}$), $19.8\, \rm{arcsec}$ ($1.6\, \rm{kpc}$), and $22\, \rm{arcsec}$ ($1.7\, \rm{kpc}$), identified as knots A, B, C, and G in \cite{2016ApJ...832....3A}, respectively. In Table\,\ref{tab_pol} we show the \textit{V} and \textit{I}~band polarisation degree and position angle measured for each knot with NOT/ALFOSC, compared with the \textit{HST}/F606W measurements by \citet{2016ApJ...832....3A}. Overall there is a good agreement between ALFOSC measurements in the \textit{V}~band and the F606W filter in \textit{HST}. The extended polarisation signal over $\gtrsim 0.6\, \rm{arcsec}$ scales is well recovered by ALFOSC, although compact ($\sim 0.1\, \rm{arcsec}$) shock-like features reaching $\sim 60$ per cent polarisation are diluted in our images due to the larger seeing. This is also the case for knots D, E, F, and I ($2$--$4$ per cent), as they are located at the narrowest part of the jet. The overall emission of knots A, B, C, and G is spatially resolved by ALFOSC. Note that the data from \citet{2016ApJ...832....3A} were obtained $\gtrsim 10$ years before our observations, thus possible variability might be also present in the knots. The values of the linear polarisation degree reported here would drop by a factor $\sim 2$ if the extended starlight from the host galaxy would not have been subtracted (see Section\,\ref{obs}). The extended starlight image contains only the lowest spatial frequencies, excluding the frequencies associated with the size of the extended knots A, B, C, and G. This was confirmed after inspecting the light profiles in the reconstructed extended image.

Regarding the circular polarisation, no signal was detected above a $3\sigma$ level of \textit{P}$_{\rm circ} \simeq 1.5$ per cent in either the \textit{V} or the \textit{I}~bands, suggesting that there is no conversion of linear into circular polarisation in the nucleus of M87. The possible explanation for the low degree of optical linear polarisation in the nucleus of this galaxy is discussed in Section.\,\ref{discuss}
\begin{table*}
 \begin{center}
   \caption{NOT/ALFOSC measurements of the linear polarisation degree (\textit{P}$_{\rm lin}$) and its position angle ($\chi$) in the \textit{V} and \textit{I}~bands for the different knots identified along the jet of M87. These are compared with \textit{HST} measurements by \citet{2016ApJ...832....3A} and the variability study by \citet{2011ApJ...743..119P} for the nucleus and HST-1.}\label{tab_pol}
\begin{threeparttable}[b]
 \begin{tabular}{cccccccc}
 \hline
 \bf Knot name & \bf Distance & \bf \textit{P}$_{\rm lin}(V)$ & \bf  $\chi_V$ & \bf \textit{P}$_{\rm lin}(I)$ & \bf $\chi_I$ & \bf \textit{P}$_{\rm lin}(\rm{F606W})$ & \bf $\chi_{\rm F606W}$ \\
  & (kpc) & (per cent) & (deg) & (per cent) & (deg) & (per cent) & (deg) \\[0.5ex] 
 \hline
 Nucleus & -- & $3.1 \pm 0.6$ & $44 \pm 6$ & $2.3 \pm 0.4$ & $50 \pm 7$ & $1.0$--$10.7$\tnote{a} & --\tnote{b} \\
 HST-1+D+E+F+I & $0.06$--$0.87$ & $2$--$4 \pm 0.7$ & $90 \pm 20$ & $1.3$--$4.0 \pm 0.5$ & $98 \pm 14$ & $20$--$45$\tnote{c} & $142$--$165$ \\
 A & $0.87$--$1.04$ & $9$--$20 \pm 1$ & $50 \pm 6$ & $4.9$--$14.7 \pm 0.7$ & $54 \pm 7$ & $20 \pm 1$ & $19 \pm 3$ \\
 B & $1.07$--$1.31$ & $7$--$20 \pm 1$ & $129 \pm 10$ & $3.6$--$13.1 \pm 0.7$ & $130 \pm 10$ & $16$--$23$ & $166$--$190$ \\
 C & $1.36$--$1.52$ & $6$--$9 \pm 1$ & $36 \pm 10$ & $2.5$--$8.0 \pm 0.8$ & $36 \pm 7$ & $8$--$16$ & $147$--$171$ \\
 G & $1.54$--$1.80$ & $7$--$17 \pm 2$ & $55 \pm 13$ & $2.4$--$7.9 \pm 0.8$ & $56 \pm 9$ & $22$--$27$ & $7$--$18$\\
 \hline
\end{tabular}
\begin{tablenotes}
  \item[a] The range given in polarisation degree corresponds to the values shown by the core during the 2002--2007 epoch.
  \item[b] The nuclear polarisation angle seen by \textit{HST} is not reproduced here due to the high and wide variability that covers essentially any possible angle.
  \item[c] The range given refers to the spatial variation between HST-1 and the other knots, but HST-1 itself shows variability in this interval \citep{2011ApJ...743..119P}.
\end{tablenotes}
\end{threeparttable}
\end{center}
\end{table*} 

\begin{figure}
\centering
\includegraphics[width=\columnwidth]{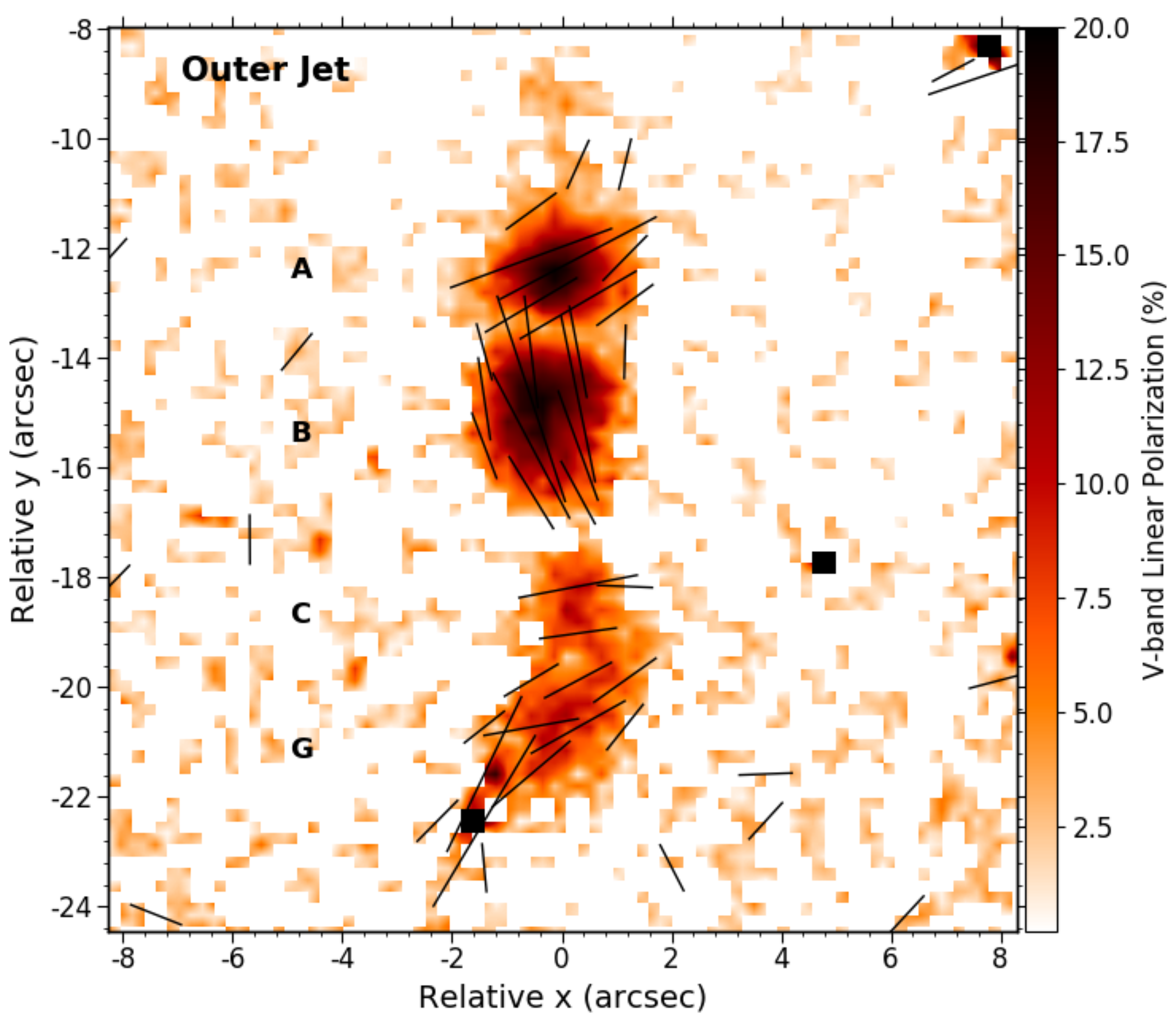}
\includegraphics[width=\columnwidth]{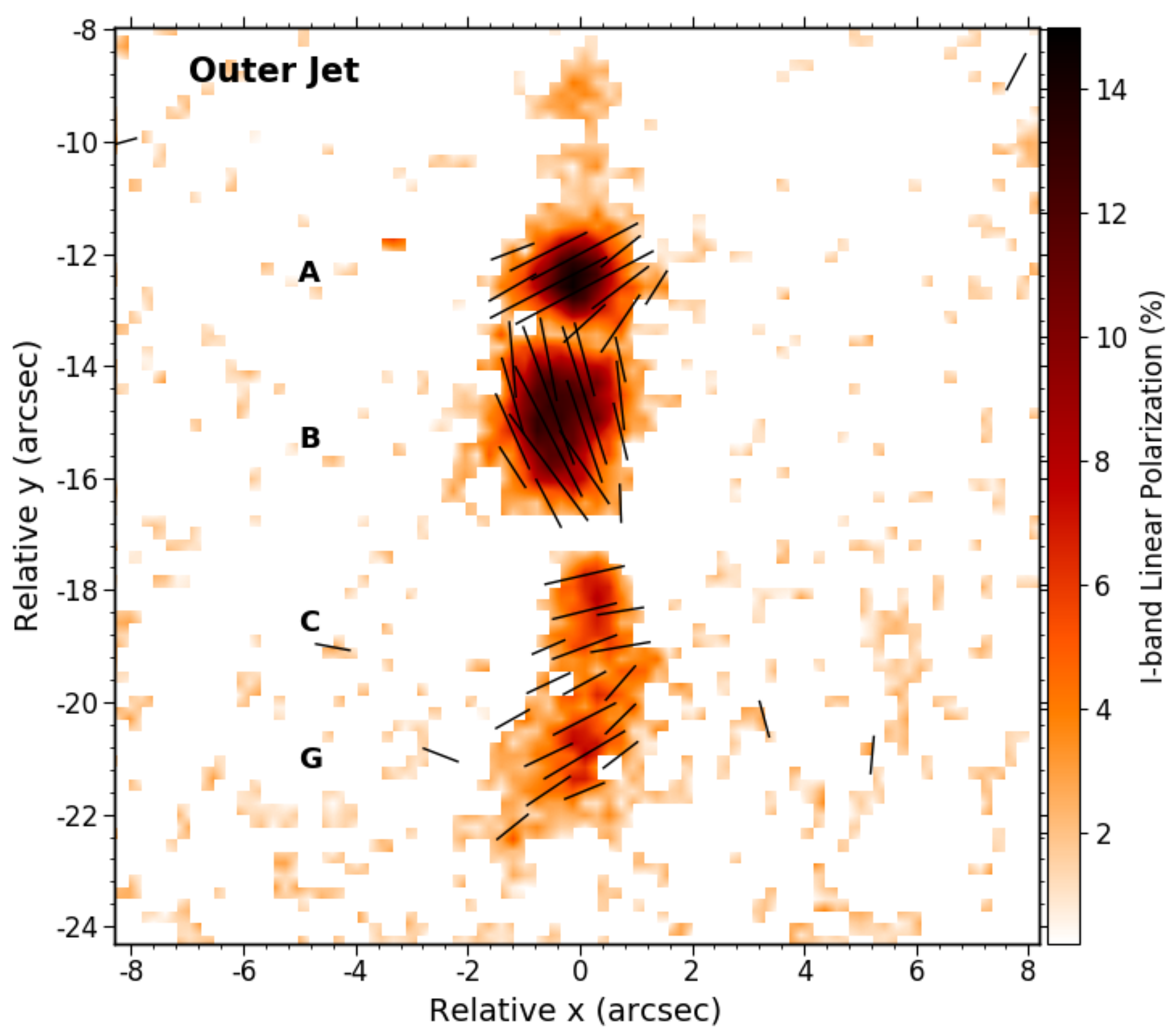}
\caption{Zoom of Fig.\,\ref{fig_pa} for the farthest jet region including the knots A, B, C, and G. The upper and lower panels correspond to the \textit{V} and the \textit{I}~bands, respectively.}\label{fig_zoom}
\end{figure}

The angle of the linear polarisation shows a remarkable helical structure along the jet axis, as described by \citet{2016ApJ...832....3A}. It is oriented nearly perpendicular to the jet axis orientation at the nuclear position (see Fig.\,\ref{fig_pa}), with \textit{PA}\,$= 44\degr \pm 6\degr$ ($50\degr \pm 7\degr$) in the \textit{V}~band (\textit{I}~band). This differs from the value reported by \citet{2016ApJ...832....3A}, although the linear polarisation angle distribution shown in their fig.\,3 (lower-right panel) seems to be in fair agreement with our map in Fig.\,\ref{fig_pa}. At the core the polarisation angle shows a high variability in very short timescales, with significant changes within weeks \citep{2011ApJ...743..119P}. At farther distance the polarisation angle shifts by $45\degr$ towards values in the $70\degr$ to $110\degr$ range ($84\degr$ to $112\degr$ \textit{I}~band) within the inner knots D, E, F, and I. A zoomed of the outer jet is shown in Fig.\,\ref{fig_zoom}. Knot A shows a polarisation angle which is again in phase with the nucleus, $50\degr \pm 6\degr$ ($54\degr \pm 7\degr$), followed by a nearly perpendicular orientation in knot B with $129\degr \pm 10\degr$ ($130\degr \pm 10\degr$). In knot C another abrupt change aligns the linear polarisation completely perpendicular to the jet axis at $36\degr \pm 10\degr$ ($36\degr \pm 7\degr$). Finally, knot G returns to the nuclear alignment with $55\degr \pm 13\degr$ ($56\degr \pm 9\degr$). Note the sharp drop in the degree of linear polarisation at the interface between knots with different \textit{PA} orientation, reaching values close to zero. The resulting distribution of \textit{PA} is consistent, within the errors, with the maps obtained if no extended starlight would have been subtracted before the polarimetric analysis (Section\,\ref{obs}), which means that there are not significant residuals introduced in this step.


\section{Discussion}\label{discuss}

M87 became a famous target early since the discovery of its optical jet, and has been since then the subject of numerous studies in the literature. In this Section we briefly put our results in the context of previous works, to focus then on two of the most important aspects of the jet revealed by the polarimetry: the helical jet structure and the low degree of optical polarisation in the nucleus.

\subsection{Previous works}
The presence of strongly polarised optical emission ($\gtrsim 30$ per cent) associated with synchrotron radiation in the jet of M87 was confirmed early in the 1950's by \citet{1956ApJ...123..550B}, shortly after the identification of M87 as the optical counterpart of a strong radio emitter. In this pioneering study the authors already noted the change in the linear polarisation vector between the knots A--C and the knot B. Further observations by \citet{1959ApJ...130..340H} and \citet{1978ApJ...220L..31S} provided the first measurements of the linear polarisation degree for the most prominent knots in the jet. Later, the jet core polarisation was marginally detected in the UV with a value of $3 \pm 2$ per cent within the inner $1.4\, \rm{arcsec}$ by \citet{1989A&A...224...17F}.

The high-angular resolution observations and multi-epoch campaigns in the optical with \textit{HST} revealed the small scale structure of the jet \citep{1999AJ....117.2185P} and the temporal variability of the core and knots, including the extraordinary flare of HST-1 ($2005$--$2006$; \citealt{2009ApJ...699..305H,2011ApJ...743..119P}). The core shows a lower degree of optical linear polarisation when compared to the knots, although its amplitude can change by an order of magnitude ($\sim 1$--$10$ per cent; \citealt{2011ApJ...743..119P}). During our observations in 2017 the nucleus of M87 was in a relatively quiet state, especially HST-1 which is not detected in our images despite the angular resolution achieved by NOT/ALFOSC during the night ($0.6\, \rm{arcsec}$), and fainted almost completely in the \textit{Chandra} X-ray observation about a month earlier \citep{2019ApJ...879....8S}. 

For the circular polarisation in the optical range, there is only one previous measurement of M87 \citep{1972ApJ...177..177K} but the nucleus was not detected. Submillimetre interferometric observations at $230\, \rm{GHz}$ by \citet{2014ApJ...783L..33K} were also not successful in detecting the core, as shown by their fig.\,1. However, this frequency is very close to the jet turnover \citep{2015ApJ...814..139K,2016MNRAS.457.3801P}, where the circular polarisation degree of synchrotron self-absorbed emission is expected to cancel out \citep{1970ranp.book.....P,1977OISNP..89.....P}. Regarding the present work, we did not detect circular polarisation in the optical above an upper limit of \textit{P}$_{\rm circ} \leq 1.5$ per cent in the nuclear region of M87.

\subsection{The polarisation pattern and the helical jet}
Helical magnetic fields are tightly associated with the launch and propagation of jets in AGN \citep{2011AstL...37..154M}, as they develop efficiently due to the magnetised rotation of the SMBH and the accretion disc \citep{2001NewA....6...61N,2016MNRAS.461L..46T}. In this scenario, the ejected plasma from the inner region/corona are accelerated along the jet axis, tracing helical trajectories that follow the magnetic field lines. The emitted radiation is linearly polarised and --\,in the case of synchrotron self-absorbed radiation\,-- the polarisation vector of the optically-thin radiation in the jet frame is perpendicular to the magnetic field distribution, but it is parallel to the magnetic field in the optically-thick part at lower frequencies \citep{1970ranp.book.....P,1977OISNP..89.....P}.

However, the polarisation degree and its orientation in the observer's frame is strongly modulated by relativistic beaming effects, typically producing a bimodal distribution of positions angles which are either parallel or orthogonal to the jet axis \citep{2005MNRAS.360..869L}. In M87 this configuration is clearly seen in the knots distribution along the jet, where the \textit{PA} changes by $\sim 90\degr$ between a knot and its neighbour (Fig.\,\ref{fig_pa}), as it is also seen in the \textit{HST} polarisation maps by \citet{2016ApJ...832....3A}, summarised in Table\,\ref{tab_pol}. A similar behaviour has been seen during a bright flare event in the jet of the blazar BL Lacertae \citep{2008Natur.452..966M}. The propagation of a shock wave in this nucleus is accompanied by drastic $\sim 90\degr$ changes in the \textit{PA} of the optical linear polarisation, possibly tracing the cycling of the ejected component along the helical structure of the magnetic field \citep{2005MNRAS.360..869L}. Furthermore, the degree of polarisation drops abruptly in the transition from the parallel to the orthogonal configuration of the linear polarisation angle, caused by the mix of polarised radiation from two regions with different magnetic field orientations \citep{1993Ap&SS.206...55S}. The temporal behaviour of a single jet component in BL Lacertae is remarkably similar to the polarisation pattern in the knots of M87 jet.

The comparison of the optical polarisation maps with that at $22\, \rm{GHz}$ by \citet{2016ApJ...832....3A} reveals a general agreement in the position angle of the linear polarisation between radio and optical data, except for two regions: the nucleus and HST-1. In HST-1 the optical maps show an orthogonal distribution with respect to that in radio, where the polarisation angle is parallel to the jet axis. A similar result was obtained earlier by \citet{1999AJ....117.2185P} from the comparison of \textit{HST} observations and a VLA map at $15\, \rm{GHz}$. This distribution is expected if the synchrotron radiation in the optical range is optically-thin and the radio emission is optically-thick \citep{1970ranp.book.....P,2014ApJ...780...87M}. The comparison is not so straightforward in the case of the nucleus, since the polarisation angle covered essentially all the possible orientations over a six years period \citep{2011ApJ...743..119P}. However, the spectral energy distribution at the core reveals a flat radio spectrum with a turnover at $\sim 400\,\rm{GHz}$ (\mbox{\citealt{2015ApJ...814..139K}},\mbox{\citealt{2016MNRAS.457.3801P}}), leading to an inverted power law at higher frequencies (\mbox{\citealt{2001ApJ...551..206P}},\mbox{\citealt{1997MNRAS.285..181S}}) characteristic of optically-thin synchrotron radiation \citep{1970ranp.book.....P}. This suggests that the nuclear emission is optically thick at $22\, \rm{GHz}$, in line with the lack of correlation found among the radio-to-optical spectral indices in the knots by \mbox{\citet{2001ApJ...551..206P}}. Future EHT polarimetric observations at the highest spatial resolution close to the turnover frequency can probe this scenario. Specifically, a steep increase of the circular and linear polarisation degrees is expected with increasing frequency above the turnover.

The linear polarisation maps reveal a sharp increase in both the total flux and the linear polarisation degree at the position of knot A ($\sim 10\, \rm{arcsec}$; $0.8\, \rm{kpc}$ projected distance from the nucleus), possibly associated with a shock and a recollimation in the jet. This event may be associated with a change in the ISM conditions at a given distance from the nucleus. For instance, filaments of ionised gas are present in the central few hundred parsecs \citep{1994ApJ...435L..35H}. Moreover, \citet{2018MNRAS.474.4169O} inferred a change in the stellar initial mass function within the innermost $0.5\, \rm{kpc}$, which could be caused by a denser ISM in the central region.

No circular polarisation is detected in any of the jet knots. According to \citet{1977OISNP..89.....P}, the highest degree of circular polarisation is expected at frequencies slightly higher than the jet turnover ($\gtrsim 400\, \rm{GHz}$ for the core of M87), decreasing as \textit{P}$_{\rm circ} \propto \nu^{-1/2}$ with increasing frequency. If the knots turnovers are located at a similar frequencies as the core, the non detection of circular polarisation could be explained by the expected decay and the lack of conversion from linear polarisation. This scenario could be explored using future interferometric observations with ALMA to probe the polarised light in M87 at frequencies above the jet turnover ($\gtrsim 400\, \rm{GHz}$) at parsecs scale resolution.

\subsection{Low optical nuclear polarisation}
The distinctive characteristic in the polarisation maps of M87 (Fig.\,\ref{fig_polmap}) is the lower degree of optical linear polarisation in the core, when compared to the high values in the jet ($\gtrsim 10$--$40$ per cent). This difference cannot be ascribed to variability: the nucleus hardly ever exceeds $10$ per cent fractional polarisation \citep{2011ApJ...743..119P}. Even during the high activity period in 2005--2006, the difference between the nucleus ($\lesssim 12$ per cent) and HST-1 ($20$--$40$ per cent) was remarkable. The present observations further show that the low degree of optical linear polarisation in the core is not caused by a conversion of linear into circular polarisation, within the upper limit estimate (\textit{P}$_{\rm circ} < 1.5$ per cent), as it should be expected in the optically-thin region. On the other hand, at radio-to-millimetre wavelengths the core is expected to show a low polarisation degree, since the jet is optically thick and additional depolarisation mechanisms become efficient in this range \citep{2017ApJ...843L..31B}.

A possible explanation to the low optical degree of polarisation could be ascribed to the magnetic field configuration. Disordered field lines at the core would be consistent with the low degree of linear and circular polarisation. Still, the sharp transition from a disordered magnetic field at the core to a highly ordered field in HST-1 \citep{2016ApJ...832....3A} should be explained in this scenario. Alternatively, the core optically-thick emission has a known complex structure with a number of jet components or knots (e.g. \citep{2018ApJ...855..128W,2020A&A...637L...6K}. This is likely also the case of the optically thin emission, thus the linear polarisation emission may include several knots with different position angles. In this case the magnetic field within each knot could be ordered, but the different position angles of the individual components would result in the effective depolarisation of the integrated emission. This is in agreement with the large spread observed in the nuclear polarisation angle, which spans almost every possible orientation over a few years timescale \citep{2011ApJ...743..119P}. In this context, the large dispersion in \textit{PA} could be associated with the flux variability of these unresolved knots, whose alternating dominance within the unresolved beam would show an erratic behaviour in the integrated \textit{PA}. This could be similar to the flickering activity in Sgr\,A$^*$ in the near-IR, which is associated with continuous changes in the \textit{PA} of the linear polarisation as recently found by GRAVITY  \citep{2018A&A...618L..10G}. A complex nuclear structure in M87 was also suggested by the flux distribution modelling by \citet{1997MNRAS.285..181S}, including radio, IR, and optical measurements at subarcsec resolution. The diminishing scale-length of the jet knots with decreasing distance from the core could be associated with the shrinking jet section and the wide opening angle at the jet base shown by \citet{2016ApJ...817..131H} at millimetre wavelengths. The further lateral expansion of the jet would allow the knots to increase in size, and therefore to exhibit a higher degree of linear polarisation when observed at higher angular resolution.

Relatively low values of the nuclear polarisation are not exclusive of M87. For instance, a similar scenario was reported for the case of M84, where no polarisation was detected in the optical range above a $3\sigma$ value of $< 8$ per cent \citep{2018ApJ...860....9M}. Or 3C\,273 \citep{1993ApJ...409..604S}, where the extremely low levels of polarisation in the optical and UV had a dramatic change in values measured only one month apart.


\section{Summary}\label{sum}

Polarimetric observations in the \textit{V} and \textit{I}~bands collected with NOT/ALFOSC in Roque de los Muchachos Observatory (La Palma, Spain) allow us to derive for the first time the four Stokes parameters in the central few kpc of M87. These observations were acquired two days before the Event Horizon Telescope campaign in early April 2017. Our main aim is to investigate the low degree of optical polarisation in the core of M87, which is optically-thin in this range, and the possible presence of circular polarisation. For this purpose we have developed the methodology needed to recover the four Stokes parameters using a $\lambda / 4$ wave-plate polarimeter, the whole set of equations is explained in detail in Section\,\ref{method}.

The linear polarisation degree of the most prominent knots in the jet (namely A, B, C, and G) reached values of about $10$--$20$ per cent, in line with previous works. The jet core shows a $2$--$3$ per cent degree of linear polarisation in the optical, consistent with a low state of nuclear activity. The latter is also supported by the non detection of HST-1 in the intensity images and the decrease of the X-ray flux from this knot by 73 per cent during the same epoch. No circular polarisation was detected above a $3 \sigma$ level of $1.5$ per cent, suggesting that the low linear polarisation in the optical range is not due to conversion into circular by any possible mechanism. Such a low degree of polarisation could be explained by the smaller scale-length of the jet knots in the core, possibly caused by the shrinking section of jet in the innermost region and our unavailability to spatially resolve their polarisation individually. Furthermore, a disordered magnetic field configuration in the innermost $10\, \rm{pc}$ could also contribute to the low degree of linear polarisation observed. Future interferometric observations with ALMA at frequencies above the jet turnover ($\gtrsim 400\, \rm{GHz}$) could probe the innermost structure of the core in the optically-thin regime and map the polarised light at parsec scales.


\section*{Acknowledgements}
We thank the referee Prof. E. Perlman for his precise comments that improved significantly the quality of this work. AYF acknowledges the national program of scholarships for postgraduate studies abroad from the Paraguayan Government ``Don Carlos Antonio L\'opez'', which allowed her to develop this research as a product of a masters program. JAFO and MAP acknowledge financial support from the Spanish Ministry of Economy and Competitiveness (MINECO) under grant number MEC-AYA2015-53753-P. JAFO acknowledges financial support by the Agenzia Spaziale Italiana (ASI) under the research contract 2018-31-HH.0.

The data presented here were obtained with ALFOSC, which is provided by the Instituto de Astrof\'isica de Andaluc\'ia (IAA) under a joint agreement with the University of Copenhagen and NOTSA. We would like to thank the local staff at the NOT telescope for their careful assistance and willingness during the preparation, observation, and further processing of the ALFOSC data. This research made use of \textsc{Astropy}, a community-developed core Python package for Astronomy (Astropy Collaboration, 2013). This research has made use of the NASA/IPAC Infrared Science Archive, which is operated by the Jet Propulsion Laboratory, California Institute of Technology, under contract with the National Aeronautics and Space Administration.

Special thanks to Celine Peroux for useful insights, to Felipe Alves and Elena Redaelli for sharing their polarisation plotting expertise.

\bsp	
\label{lastpage}
\end{document}